\documentclass[conferenc,twoside,web]{IEEEtran}
\usepackage[bookmarks=false]{hyperref}       

\usepackage{url}            
\usepackage{booktabs}       
\usepackage{nicefrac}       
\usepackage{microtype}      
\usepackage{xcolor}         

\def\journalname{Preprint. Accepted in IEEE Transactions on Medical Imaging}
\usepackage{cite}
\usepackage{amsmath,amssymb,amsfonts}
\usepackage{algorithmic}
\usepackage{graphicx}
\usepackage{textcomp}
\usepackage[utf8]{inputenc}
\usepackage{subcaption}

\newcommand{\A}{\textbf{L}}
\newcommand{\pA}{{\textbf{L}^\dag}}
\newcommand{\x}{\textbf{x}}
\newcommand{\y}{\textbf{y}}
\newcommand{\noise}{\boldsymbol{\epsilon}}

\newcommand{\mymethod}{3D-PIUNet}

\def\BibTeX{{\rm B\kern-.05em{\sc i\kern-.025em b}\kern-.08em
    T\kern-.1667em\lower.7ex\hbox{E}\kern-.125emX}}
\begin{document}
\bstctlcite{IEEEexample:BSTcontrol}
\title{Enhancing Brain Source Reconstruction by Initializing 3D Neural Networks with Physical Inverse Solutions}
\author{Marco Morik\textsuperscript{*,1,2}, Ali Hashemi\textsuperscript{1,2}, Klaus-Robert Müller\textsuperscript{1,2,3,4}, Stefan Haufe\textsuperscript{5,6,7} and Shinichi Nakajima\textsuperscript{*,1,2,8}
\thanks{This work was funded by the German Federal Ministry of Education and Research under the grant BIFOLD24B.}

\thanks{* Corresponding authors: m.morik@tu-berlin.de, nakajima@tu-berlin.de  \\
1:  The Berlin Institute for the Foundations of Learning and Data, Berlin, Germany  \\
    2:Machine Learning Group, Department of Computer Science, TU Berlin, Berlin, Germany\\
    3: Department of Artificial Intelligence, Korea University, Seoul 136-713, Korea \\
    4: Max Planck Institut für Informatik, Saarbrücken, Germany\\
    5: Bernstein Center for Computational Neuroscience, Berlin, Germany \\
  6: Physikalisch-Technische Bundesanstalt, Berlin, Germany \\
  7: Berlin Center for Advanced Neuroimaging, Charité–Universitätsmedizin Berlin, Germany \\
  8: RIKEN AIP, Tokyo, Japan.
  }
}
\maketitle

\begin{abstract}
Reconstructing brain sources is a fundamental challenge in neuroscience, crucial for understanding brain function and dysfunction. Electroencephalography (EEG) signals have a high temporal resolution. However, identifying the correct spatial location of brain sources from these signals remains difficult due to the ill-posed structure of the problem.
Traditional methods predominantly rely on manually crafted priors, missing the flexibility of data-driven learning, while recent deep learning approaches focus on end-to-end learning, typically using the physical information of the forward model only for generating training data.
We propose the novel hybrid method \mymethod~for EEG source localization that effectively integrates the strengths of traditional and deep learning techniques. \mymethod~starts from an initial physics-informed estimate by using the pseudo inverse to map from measurements to source space. Secondly, by viewing the brain as a 3D volume, we use a 3D convolutional U-Net to capture spatial dependencies and refine the solution according to the learned data prior. 
Training the model relies on simulated pseudo-realistic brain source data, covering different source distributions. Trained on this data, our model significantly improves spatial accuracy, demonstrating superior performance over both traditional and end-to-end data-driven methods.
Additionally, we validate our findings with real EEG data from a visual task, where \mymethod~successfully identifies the visual cortex and reconstructs the expected temporal behavior, thereby showcasing its practical applicability. 
\end{abstract}


\section{Introduction}
\begin{figure*}
    \centering
    \includegraphics[width=0.95\linewidth]{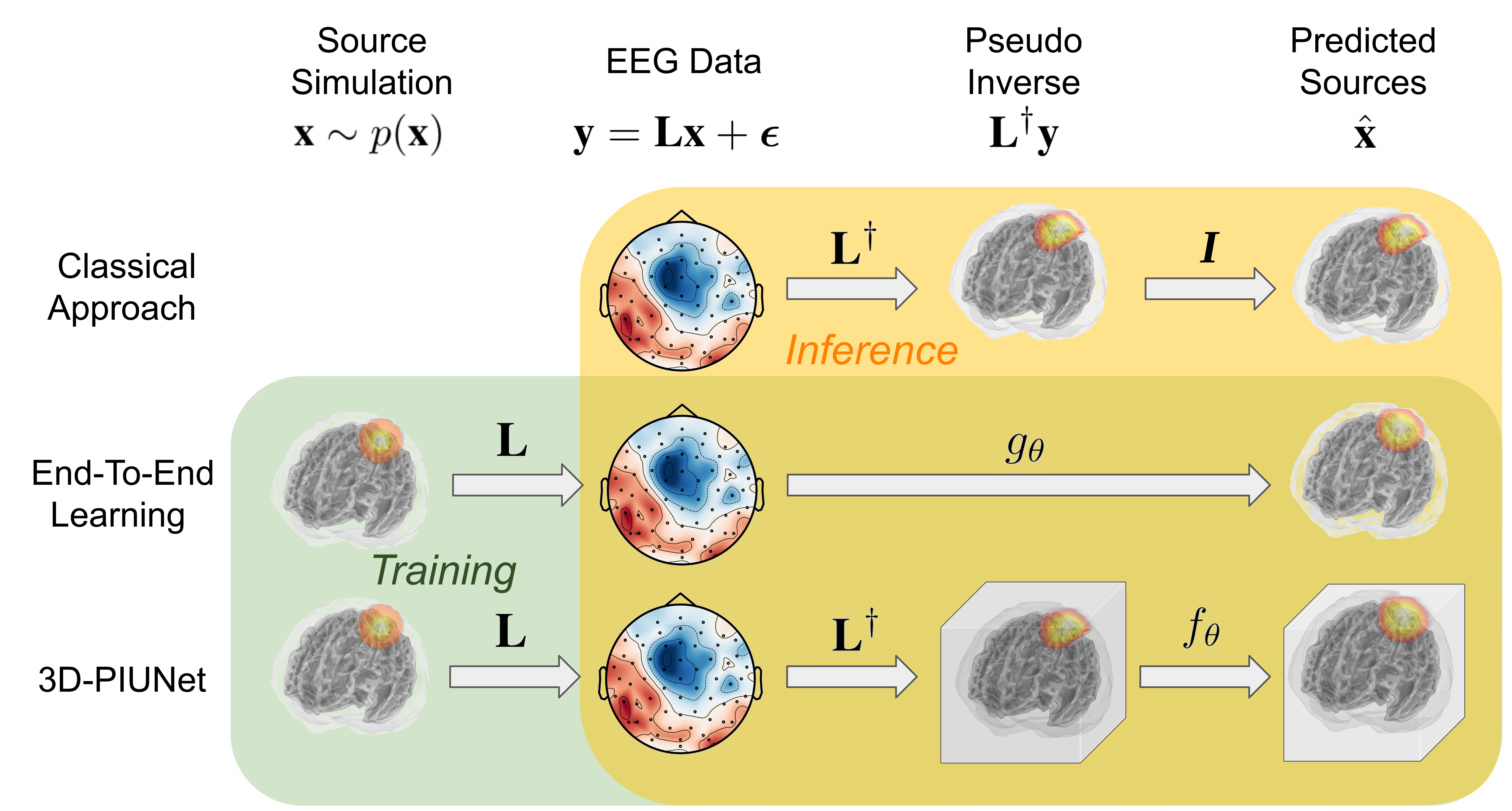}
    \caption{Overview of different approaches for source localization from EEG data. \textbf{Top} The classical (Minimum Norm) approaches uses the pseudo-inverse $\pA$ of the forward model, combined with an identity mapping $\mathbf{I}$ for illustrative purposes, to estimate the sources $\hat{\x}$ from the EEG measurements $\y$. \textbf{Middle:} Data-driven end-to-end approaches train some neural network $g_\theta$ on simulated data, where the source activity $\x$ is drawn from the data prior $p(\x)$. During inference, the network predicts the sources $\hat{\x}$ directly from measurements without explicitly using the forward model. \textbf{Bottom:} Our approach, \mymethod, integrates data-driven learning with physics-informed modeling by incorporating the forward model's physical information through the pseudo-inverse. After transforming the EEG measurements into a grid-like source space, a 3D convolutional U-Net $f_\theta$ refines the pseudo-inverse solution according to the learned data prior, resulting in an improved source estimate $\hat{x}$.}
    \label{fig:Figure1}
\end{figure*}
To understand both normal brain function and dysfunction, it is essential to accurately identify the precise sources of electrical activity within the brain, a process known as \emph{source localization}\cite{michel2004eeg}. This is an important tool in cognitive neuroscience, enabling the study of brain responses to various stimuli and the functional organization of the brain. 
Electroencephalography (EEG) and magnetoencephalography (MEG) are non-invasive techniques that allow for monitoring brain activity without the need for surgical intervention \cite{he2018electrophysiological}. Moreover, M/EEG offers high temporal resolution, capturing neural events on the order of milliseconds, which is critical for studying fast neural dynamics and the timing of cognitive processes. This high temporal resolution makes EEG particularly suitable for investigating the rapid neural processes underlying cognition, perception, and action\cite{gevins1995mapping,burle2015spatial}. 

However, pinpointing the precise brain regions responsible for these signals remains a challenge due to the ill-posed nature of the EEG inverse problem \cite{baillet2001electromagnetic}. 
Conventional methods try to address this by posing priors on the source activation, like sparsity or minimal norm activity \cite{hamalainen1994interpreting}.
Manual crafted priors bias the solution to specific settings. For example, sparse methods do not work on distributed sources and minimum norm methods do not work for sparse sources \cite{ding2008sparse}.
In recent years, neural networks \cite{delatolas2023eeg, sun2022deep,hecker2021convdip,thomas2019analyzing} have shown improved performance by learning a direct mapping from the measurement to sources in an end-to-end fashion. Unlike traditional techniques, neural networks offer greater flexibility in formulating priors on brain activation by simulated data, that serves as a proxy for real-world conditions, possibly capturing a wide range of brain activity patterns\cite{zhang2024dct}.
However, these approaches typically incorporate the forward model information only indirectly through their training data, which can limit their accuracy and generalizability in different contexts \cite{sun2023deep}.

To combine worlds, we propose a novel hybrid approach, named \mymethod, that combines the strengths of both classical methods, which provide physics-informed solutions that generalize across different forward models, and deep learning methods, which learn complex, data-driven priors to improve accuracy. We incorporate physics-informed prior knowledge from the forward model via the pseudo-inverse solution provided by eLORETA  \cite{pascual2007discrete}.
Our method leverages a 3D convolutional neural network to refine this pseudo-inverse solution according to a learned data-driven prior that is able to capture complex patterns in the source space. We show a comparison of classical approaches, end-to-end learning and our proposed \mymethod~in Fig. \ref{fig:Figure1}.

Our experimental results on a comprehensive synthetic dataset demonstrate that \mymethod~consistently outperforms both eLORETA and end-to-end deep learning methods in terms of source localization accuracy without requiring temporal information. The network effectively learns to refine the pseudo-inverse solution, leading to more accurate and robust reconstructions, especially in the presence of noise. This work highlights the potential of best-of-both-worlds approaches to overcome the limitations of traditional and data-driven end-to-end methods in EEG source imaging.

To summarize our contribution, we
\begin{itemize}
    \item propose the hybrid approach \mymethod~that includes the physics information of the forward model into the deep learning framework via its pseudo inverse,
    \item demonstrate the benefits of 3D convolution in source space to learn the spatial patterns, and
    \item extensively benchmark our approach in diverse settings, demonstrating its superiority over both classical methods and end-to-end approaches across different signal-to-noise levels, number, and size of sources.
\end{itemize}

\section{Background and Related Work}

\subsection{Problem Setup}
EEG source localization aims to identify the spatial origins of electrical activity within the brain from scalp-recorded EEG signals. The mapping from brain sources to EEG measurements can be described by a linear \emph{forward model}:
\begin{equation}
    \y = \A \x + \noise, 
\end{equation}
where a known lead field matrix $\A \in \mathbb{R}^{M\times 3N}$ maps the electrical activity of the brain, $\x \in \mathbb{R}^{3N\times T}$, to the sensor measurements, $\y \in \mathbb{R}^{M\times T}$. Here, $M$ describes the number of sensors on the scalp, $N$ the number of potential brain sources within the brain, with $3$ dimensions to allow free orientation, and $T$ the number of time steps. The \emph{noise vector} $\noise$ describes the measurement noise, covering inaccuracies in the forward model, electrical pollution and sensor errors. This work focuses on the spatial aspect of source localization and considers a single time point, i.e., $T=1$.

The goal of source localization is to invert this forward model, therefore the name \emph
{inverse problem}. As $N\gg M$, the problem is ill-posed, 
there are infinitely many possible solutions for a given set of measurements \cite{tarantola2005inverse}. Therefore, finding a unique solution requires heavy regularization. By introducing reasonable priors over the source activation, the most likely source activation can be localized more accurately \cite{michel2004eeg}. The different methods for source localization can be broadly categorized based on the assumed priors to constrain the infinite solution space.

\subsection{Classical Approaches to EEG Source Localization}
Early works on source localization assumed a limited and fixed number of sources active in the brain as a prior. In this dipole fitting both the position and orientation of the sources are adjusted to match the measurements \cite{scherg1985two,scherg1990fundamentals}. However, the number of active regions must be set in advance and the optimization problem becomes hard when trying to localize more than two dipoles. 

On the other hand, distributed methods assume that brain activity can be spread across regions involving multiple dipoles. These methods place a dense set of potential source locations within the head volume. The inverse problem becomes a regularized regression problem, predicting the amplitudes of each location, thus not constraining the number of dipoles \cite{michel2019eeg}. The Minimum Norm Estimates (MNE) \cite{hamalainen1994interpreting} predicts the distribution of sources by minimizing the $\ell_2$-norm of the source amplitudes. The $\ell_2$-norm offers a closed-form solution and is, therefore, fast to compute. However, it has a bias towards sources close to the sensors. To counteract this bias, different weighted minimum norm solutions \cite{lin2006assessing} have been proposed, such as dSPM \cite{dale2000dynamic}, LORETA \cite{pascual1994low}, sLORETA \cite{pascual2002standardized} and eLORETA \cite{pascual2007discrete} or combinations of those \cite{pellegrino2020accuracy}. As this weighting promotes a spatial smoothness prior, small sources might be missed or blurred. To obtain sparse solution with minimal non-zero sources, one can replace the $\ell_2$-norm with $\ell_1$-norm, called Least Absolute Shrinkage and Selection Operator (LASSO) \cite{tibshirani1996regression} or selective minimum norm \cite{matsuura1995selective}. By combination of sparse and dense prior, it is also possible to generate solutions with mixed spatial extent \cite{haufe2011large}.

Lastly, Bayesian frameworks integrate prior knowledge about source distributions to flexibly model brain activity \cite{tipping2001sparse}. Sparse Bayesian Learning (SBL) 
pioneered this approach by optimizing a Type-II ML cost function to enhance source estimation in noisy environments \cite{wipf2009sparse}. Building on SBL, hierarchical Bayesian inference techniques have extended this framework to simultaneously estimate both sources and noise addressing various noise structures, including homoscedastic \cite{hashemi2021unification}, heteroscedastic \cite{cai2021robust}, and full-structure \cite{hashemi2022joint} noise. 
These methods have been further extended to include coherent spatial source clusters as in cMEM \cite{chowdhury2013meg}.

\subsection{Deep Learning Methods}
In recent years, learning-based approaches have become popular in the sciences due to their flexibility and fast inference time, including EEG data analysis \cite{roy2019deep}. A better understanding of brain activity due to virtual brain models  \cite{sanz2013virtual} and large-scale dynamical models \cite{breakspear2017dynamic} allow the generation of realistic simulation data acting as a powerful data-prior. To learn this data prior, previous work focused on end-to-end learning with different network architectures. 
For single timestep predictions, fully connected networks have been used to predict source locations directly \cite{pantazis2021meg,huang2022electromagnetic} showing the same limitation in fixed number of sources as dipole fitting. Convolutional networks have been proposed to encode spatial correlation of sensor measurement \cite{hecker2021convdip}, mixing EEG and MEG signals \cite{jiao2024multi} or to incorporate the time dimension \cite{thomas2020automated}. Other recent approaches use LSTMs for temporal predictions after encoding the measurements either with residual networks \cite{sun2022deep}, graph convolutional networks \cite{jiao2022graph}, dilated convolution \cite{yu2024electrophysiological} or multi-head attention \cite{zhang2024dct}. However, all these end-to-end methods use the forward model only during data generation and require retraining or fine-tuning when the forward model changes \cite{sun2023deep}. 
Physics-informed machine learning bridges the gap and has shown superior performances in fields like fluid dynamics~\cite{raissi2019physics}, quantum chemistry~\cite{chmiela2017machine}, quantum circuits~\cite{nicoli2024physics}, and other inverse problems like Computed Tomography (CT) \cite{jin2017deep} and MRI \cite{hyun2018deep}. Dinh et al. \cite{dinh2021contextual} uses an LSTM to refine the pseudo-inverse solution by including temporal context, making a first step towards physic-inspired source localization. However, their training target was restricted to the pseudo-inverse solution, not utilizing the benefits of learning with simulation. In the following, we explain our hybrid approach \mymethod~for source localization, combining the benefits of classical and deep learning methods.

\section{Method}

Instead of learning the relationship between measurements and brain sources end-to-end, we propose a spatial-aware neural network \mymethod~to refine the eLORETA pseudo-inverse solution. This approach leverages a data prior while simplifying the problem by incorporating physical knowledge from the forward model. In the following, we first describe the benefit of starting from the pseudo-inverse, introduce the novel 3D convolutional network architecture and lastly explain the training paradigm that allows for accurate source reconstruction.

\subsection{Starting from Pseudo-Inverse Solution}
Previous works using neural networks for source estimation \cite{hecker2021convdip,sun2022deep,zhang2024dct} learn a mapping  $\hat{\x} = g_\theta(\y): \mathbb{R}^{M} \to \mathbb{R}^{3N}$ from sensor to source space. This data-driven approach allows the model to learn complex, non-linear relationships directly from simulated data, potentially capturing patterns that may be difficult to model explicitly using traditional techniques. Even though this end-to-end learning is common practice in modern deep learning, it has some drawbacks for inverse problems. Most importantly, the model needs to learn the physical dynamics only from data, while not having access to a forward model. Therefore, the learned model heavily depends on the lead fields used during training and cannot generalize well to other forward models. For example, when the number of sensors changes, classical end-to-end networks require the training of a new input layer. 

We tackle these shortcomings by starting from a pseudo-inverse solution that maps the measurements to the source space. This way, the network can leverage the established physical knowledge embedded in the forward model. Importantly, as long as the pseudo-inverse matrix maintains full column rank, which inverse methods like eLORETA achieve through regularization, this transformation preserves all information contained in the measurement vector. In cases where this condition is not met, some degradation in the reconstruction quality can occur, as the transformation loses some measurement information. During training, the model learns a data prior in source space and can focus on refining the pseudo-inverse solution according to this prior. While this framework has been applied to other inverse problems like CT \cite{jin2017deep} and MRI \cite{hyun2018deep} reconstruction, to the best of our knowledge, it is the first application to EEG data, while previous work only focused on temporal but not spatial refinement \cite{dinh2021contextual}. Given the pseudo-inverse $\pA$, the neural network is trained using the transformed measurements $\Tilde{\x} = \pA \y$ as an input. It then learns the refinement task as a mapping within source space  $f_\theta: \mathbb{R}^{3N} \to \mathbb{R}^{3N}$:
\begin{equation}
     \hat{\x} = f_\theta(\pA \y)
\end{equation}
One can interpret the pseudo-inverse solution as a non-learnable physics-informed input layer that precedes the deep neural network. This way, the pseudo-inverse can be adapted to account for different subjects at inference time, when a subject specific forward model is available. By mapping the measurements into the source space, the input dimensionality does not depend on the sensing system and the same architecture can be used, enabling transfer learning. 

Our framework is general in the sense that it works with different pseudo-inverse solutions as a starting point. While we propose the use of eLORETA, we also compared it with various pseudo-inverse solutions including MNE, sLORETA, dSPM, beamformer from the MNE package \cite{GramfortEtAl2014}, as well as the Moore-Penrose inverse, identifying $\ell_2$-norm based methods, particularly eLORETA, performing best. 
The superiority of eLORETA might be due to its zero-localization error for single sources by compensating in the computation of the pseudo-inverse matrix $\pA$ for depth and noise. 
We used a python implementation based on the MEG \& EEG - Toolbox of Hamburg\footnote{(METH,  Version: April 15, 2019) \url{https://www.uke.de/dateien/institute/neurophysiologie-und-pathophysiologie/downloads/meth_2019_04.zip}}
with 0.05 regularization strength to compute the eLORETA pseudo inverse.

\subsection{3D Convolution}
\begin{figure}
    \centering
    \vspace{-0.7cm}
    \includegraphics[width=\linewidth]{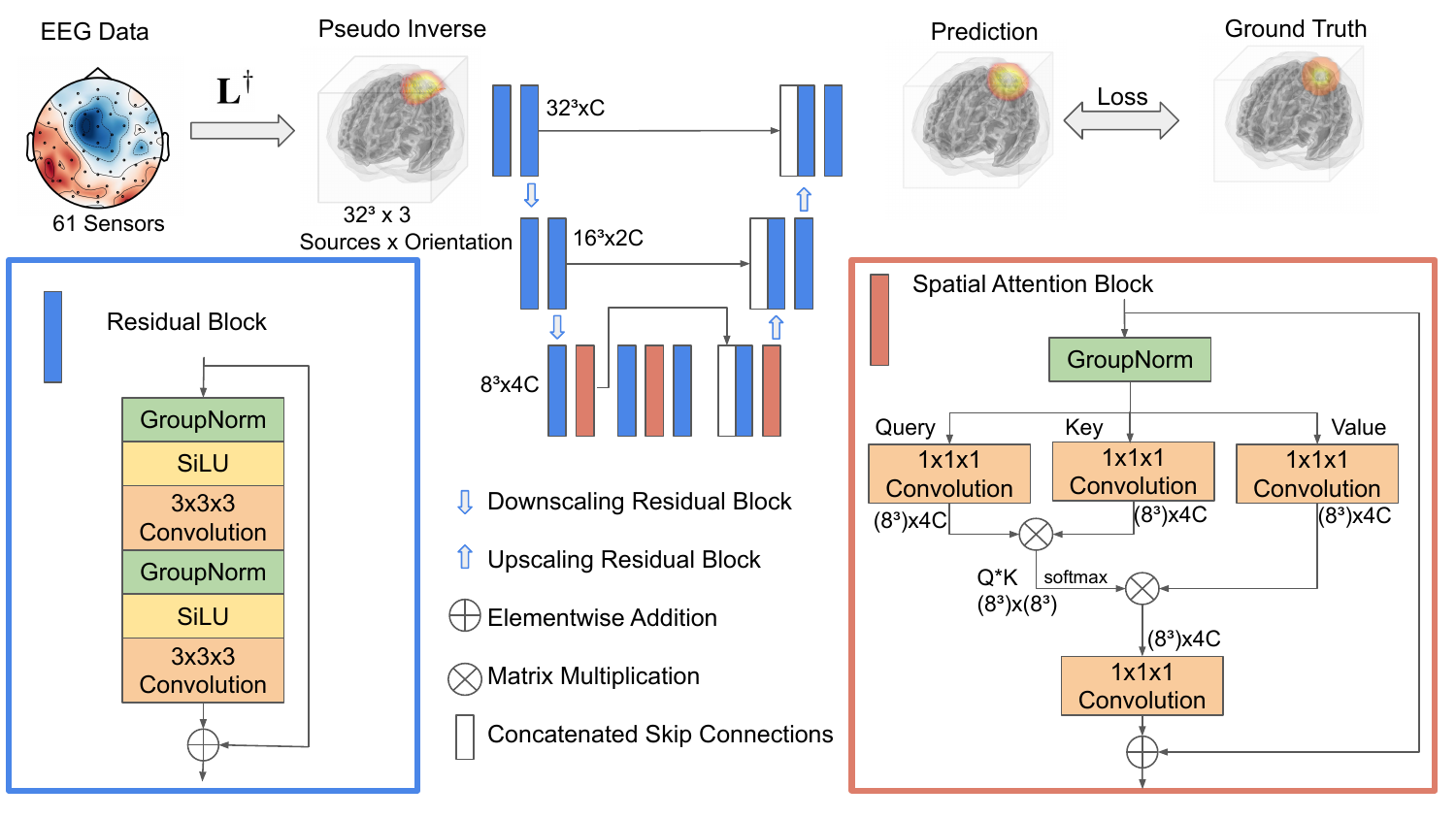}
    \caption{An overview of the \mymethod~architecture. First, the eLORETA pseudo inverse $\pA$ maps the EEG measurement to a 3D source space of $32\times32\times32$ ($32^3$) resolution. Next, a series of 3D convolutional residual blocks encodes the pseudo-inverse down to an $8\times8\times8$ latent representation. The skip connections between the encoder and decoder allow the model to preserve high-resolution information. At the latent resolution, spatial attention blocks facilitate enhanced interactions between voxels.}
    \label{fig:unetarchitecture}
\end{figure}
We divide the brain sources to a regular 3D grid. This allows the use of a 3D convolutional neural network (CNN)~\cite{lecun2015deep}.
CNNs excel in this domain due to their parameter efficiency and ability to capture spatial relationships. For tasks like source localization, where preserving spatial information across scales is crucial, the U-Net architecture~\cite{ronneberger2015u} is particularly well-suited. It employs skip connections between the encoding (downscaling) and decoding (upscaling) blocks, allowing information from the input to propagate directly to the decoder and aiding in the reconstruction of fine details. For this work, we adopt a state-of-the-art U-Net architecture\cite{nichol2021improved}. We illustrate the network architecture of \mymethod~in Fig. \ref{fig:unetarchitecture}.

Before applying the network to our pseudo sources $\Tilde{x}$, we use zero padding, extending the potential source locations to a regular $32 \times 32 \times 32$ grid. The architecture begins by lifting the pseudo inverse to a $C=32$ channel feature map, effectively creating multiple representations of the initial source estimate. Next, the network progressively down-samples, by factors of 2, to a bottleneck resolution of ($8 \times 8 \times 8$) while doubling the feature channels after each down convolution, increasing the receptive field of each convolutional layer.
At each resolution, we use two 3D convolutional residual blocks (depicted in blue). Each residual block starts with a Group Norm layer~\cite{wu2018group}, a normalization technique that is less dependent on batch size, leading to improved training stability and performance. This is followed by SiLU activation, a smooth, non-linear activation function outperforming ReLU for deep neural networks \cite{elfwing2018sigmoid}. The core computation is done by the $3\times 3 \times 3$ convolution layers that aggregate neighboring information to learn descriptive features and refine the source estimates.
Each block uses residual connections, where the input of each block is added to its output. This improves the gradient flow during backpropagation and simplifies the optimization problem \cite{he2016deep}.

After the convolutional blocks reduce the spatial dimensionality, we employ Spatial Attention Blocks (depicted in red) to enhance feature representation. These blocks use three separate linear transformations via $1 \times 1 \times 1$ convolutions that share weights across all spatial locations, serving as position-wise linear projections. The resulting tensors are then flattened along the spatial dimensions, resulting in $8^3 = 512$ tokens and producing the query $Q\in \mathbb{R}^{512 \times 4C}$, key $K\in \mathbb{R}^{512 \times 4C}$, and value $V\in \mathbb{R}^{512 \times 4C}$ matrices required for the attention mechanism~\cite{vaswani2017attention}. 
The attention operation involves computing the matrix product of the query and key, scaled by the square root of the number of channels of query, key and value $d_k=4C$, and applying a softmax function to generate the attention map. This map is used to weight the value matrix, thereby enabling each voxel to prioritize the most relevant information from other locations:
\begin{equation}
    \text{Attention}(Q,K,V) = \text{softmax}\left(\frac{Q K^\top }{\sqrt{d_k}}\right) V
\end{equation}
The resulting matrix is reshaped back into the 3D volume and is subsequently processed through a $1 \times 1 \times 1$ convolution for a final linear projection. The resulting feature maps are then added to the input of the attention block, effectively integrating contextual information from across the spatial domain. In contrast to Vision Transformers \cite{dosovitskiy2020image} that represent a group of pixels as a single token, we use a token for each voxel, but for computational reasons only in low-resolution embedding space. The attention mechanism allows the model to focus on the most relevant regions for source reconstruction. These lower layers of the network capture a broader spatial context, and attention allows the model to weigh the importance of different regions within this context, leading to more accurate refinement of the source estimates. 

Overall, the architecture incorporates a learned data prior, enabling refinement of the pseudo-inverse solution. We will show in the experiment section \ref{sec:experiments}, that this architecture improves the accuracy of source reconstruction in comparison to end-to-end models. 

\subsection{Training Details}
To train \mymethod, we explored several loss functions proposed in the literature. Classical methods, like minimum norm estimates, typically optimize the mean squared error loss in measurement space \cite{hamalainen1994interpreting}. However, this approach often overfits to noise, particularly in low SNR conditions, where noise dominates the signal. Since we train with simulated data, we have access to paired signal and source data $(y,x)$, allowing us to compute the loss in noise-free source space. While related work has used cosine similarity \cite{hecker2022long} or $\ell_2$-loss in source space \cite{sun2022deep}, we found the mean absolute error ($\ell_1$) to be a more suitable option:
\begin{align}
    \text{Loss}(\x,f_\theta(\pA \y)) = ||\x-f_\theta(\pA \y)||_{1}
\end{align}
Compared to $\ell_2$-loss, $\ell_1$-loss is less sensitive to outliers, which is crucial in source reconstruction as outliers can arise from various sources, such as artifacts or model mismatches. The $\ell_1$-loss helps prevent these outliers from disproportionately influencing the learning process. Additionally, source activity is often sparse, meaning that only a few brain regions are actively generating signals at any given time. The $\ell_1$-loss encourages the model to learn sparse solutions, aligning with this prior knowledge about brain activity.

We used the Adam optimizer  \cite{kingma2014adam} with a learning rate of 0.00001 and a batch size of 32. The training process consisted of 60000 steps, where the data is generated via the simulation explained in section \ref{sec:datageneration}. The code is available under \url{https://github.com/MarcoMorik/3D-PIUNet}.

\section{Experimental Setup}
In this section, we describe how the training and testing data for the neural network was generated and introduce the performance metrics for evaluation.
\begin{figure*}[h]
    \centering
    \includegraphics[width=\textwidth]{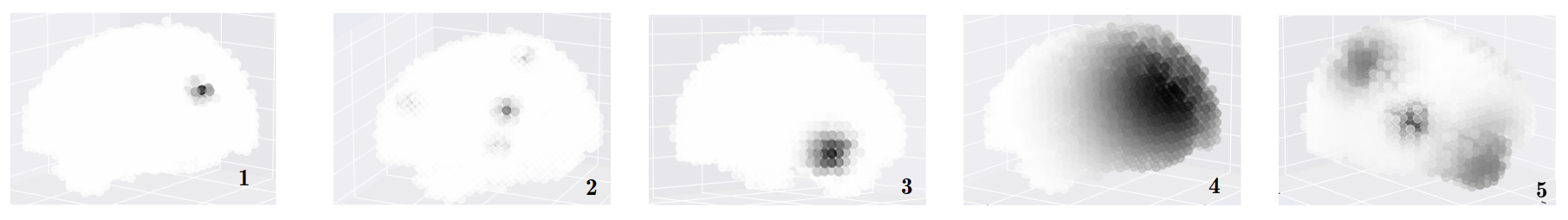}
    \caption{We generate a diverse setting of brain activation. From left to right, we have \textbf{1.} a single active source with minimal width (10mm), \textbf{2.} 4 active sources with minimal width, \textbf{3.} a single medium-sized (20mm) active source, \textbf{4.} a single large active source (80mm), and \textbf{5.} a diverse sample with 4 active sources of different size, where two sources are partly overlapping.}
    \label{fig:data-generation}
\end{figure*}

\subsection{Forward Model}
\label{sec:forwardmodeling}
To compute the forward model, we employ a three-shell Boundary Element Method (BEM) \cite{fuchs2002standardized} approach implemented within the MNE-Python library \cite{GramfortEtAl2013a}. To represent potential sources of brain activity, a volumetric source space is constructed. This source space consists of a three-dimensional grid with a uniform voxel spacing of 7 mm. For training, the fsaverage \cite{fischl1999high} head geometry is used with three concentric layers representing the scalp, skull, and brain tissues and standard conductivity values assigned to these layers: $0.3 S/m$ (scalp), $0.006 S/m$ (skull), and $0.3 S/m$ (brain). These values are based on established studies that provide realistic approximations for human head tissues \cite{hecker2021convdip}.

To calculate the forward solution, the minimum distance between potential sources and the inner skull surface is set to 5 mm. This parameter helps exclude source locations that are unrealistically close to the skull boundary, resulting in $N=4819$ sources. The volumetric source space is constructed with freely oriented sources, allowing each voxel to represent neural activity with a three-dimensional vector for direction and amplitude, providing greater flexibility in modeling complex neural activity patterns. 
Sensor positions are defined using a standard Easycap-M10 montage with $M=61$ sensors. The resulting lead field matrix $\A \in \mathbb{R}^{(61 \times (3 \cdot 4819))}$ describes the mapping from source locations and orientation to sensors.

As using the same model for both forward and inverse solutions can lead to over-optimistic results --- the so-called inverse crime \cite{colton1998inverse,kaipio2007statistical} --- we generate a second BEM solution for testing.
To account for variations in head geometry, we use a different BEM model generated from MRI data provided by the Open Access Series of Imaging Studies 1 (OASIS1)\cite{marcus2010open}. 
For test data generation, the conductivity is changed to $0.332 S/m$ (scalp), $0.0113 S/m$ (skull), and $0.332 S/m$ (brain) following an established setup \cite{hecker2021convdip,wolters2010combined}. Especially the doubling of skull conductivities has a strong influence on the resulting forward model \cite{vorwerk2019influence}. This ensures that our models are validated against a different forward model, simulating more realistic conditions and enhancing the robustness of our validation process.

\subsection{Data Generation}
\label{sec:datageneration}

Similar to previous work \cite{hecker2021convdip}, we employ a controlled synthetic EEG data generation process to facilitate model evaluation with known ground truth. To isolate the spatial reconstruction aspect of the problem, we simulate source activation at a single point in time. 
The source activations are modeled as 3-dimensional Gaussian bell curves centered at random locations $c$ within the brain\cite{haufe2011large}. The width of each activation $\sigma$ is given by the standard deviation of the bell curve. This allows us to simulate both localized and diffuse patterns of brain activity. Sources are assigned random orientations  $\Vec{v}$ and their maximum amplitudes $a\sim \mathcal{N}(0,1)$ are sampled from a standard normal distribution. For each training example, we randomly select a number of simultaneously active distributed sources $S$, sampled from a uniform distribution $\mathcal{U}(1, N_\mathrm{Activ})$, where $N_\mathrm{Activ}$ defines the maximum number of active sources allowed. For each of these sources, we sample the center $c_s$, amplitude $a_s$ and unit norm direction $\Vec{v}_s$. We calculate the activation of the source location  $\x_i$ by the sum of the $S$ active sources:
\begin{equation}
    \x_i = \sum_{s=1}^{S} a_s \frac{1}{\sqrt{2\pi\sigma_s^2}} e^{-\frac{d(p_i, c_s)^2}{2\sigma_s^2}}\Vec{v}_s
\end{equation}
Here, $d$ measures the Euclidean distance between the 3D position $p_i$ of the source location $x_i$ and the center of each active source $c_s$.
To generate the measurements $ \y = \A \x + \epsilon$, the brain activation is projected onto sensor space using the lead field matrices $\A$ described in the previous section. To introduce realistic challenges, zero-mean Gaussian noise $\epsilon \sim \mathcal{N}(0,\sigma_\mathrm{Noise}^2I)$ with varying signal-to-noise ratios (SNR), ranging from 0 dB to 30 dB, is applied to the measurements. More precisely, the noise standard deviation is given by:
\begin{equation}
 \sigma_\mathrm{Noise} = \frac{\sigma_\mathrm{Signal}}{\sqrt{10^{0.1\mathrm{SNR}}}}.
\end{equation}
This leads to the following hyperparameter that influences the generation:
\begin{itemize}
    \item \textbf{Number of Activation $N_\mathrm{Activ}$:} Ranges from 1 to 20, enabling simulation of varying source complexity.
    \item \textbf{Source Width ($\sigma$):} Spans from 10~mm to 80~mm, governing the spatial extent of each source activation, allow for simulating both diffuse and localized activity.
    \item \textbf{SNR:} Varied between $0$ dB and $30$ dB to introduce realistic noise levels and challenge the model's ability to reconstruct sources under different signal quality conditions. 
\end{itemize}
Fig. \ref{fig:data-generation} illustrates the variety of generated source patterns. While this controlled activation model allows for a systematic evaluation of spatial source reconstruction, the addition of temporal dynamics or more realistic brain activation, e.g. including brain parcellation maps\cite{destrieux2010automatic,eickhoff2018imaging}, could further improve the learned priors.

\subsection{Evaluation Metrics}

\begin{figure*}[t]
    \centering    
    \vspace{-0.2cm}
    \begin{subfigure}[b]{0.21\textwidth}
        \centering
        \includegraphics[width=\textwidth]{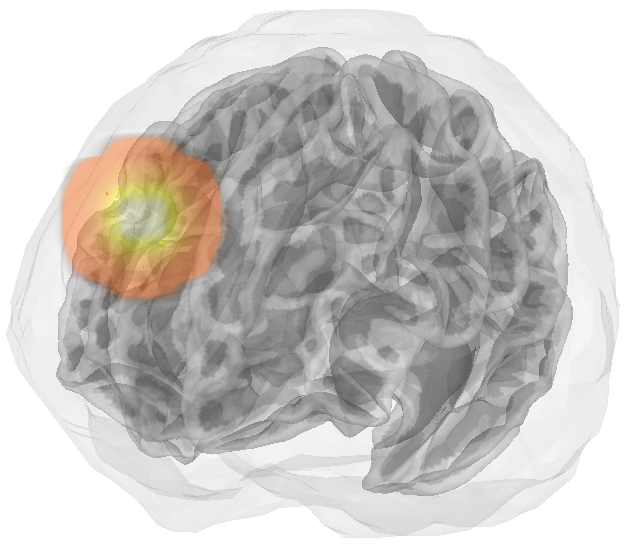}
    \end{subfigure}
    \hfill
    \begin{subfigure}[b]{0.21\textwidth}
        \centering
        \includegraphics[width=\textwidth]{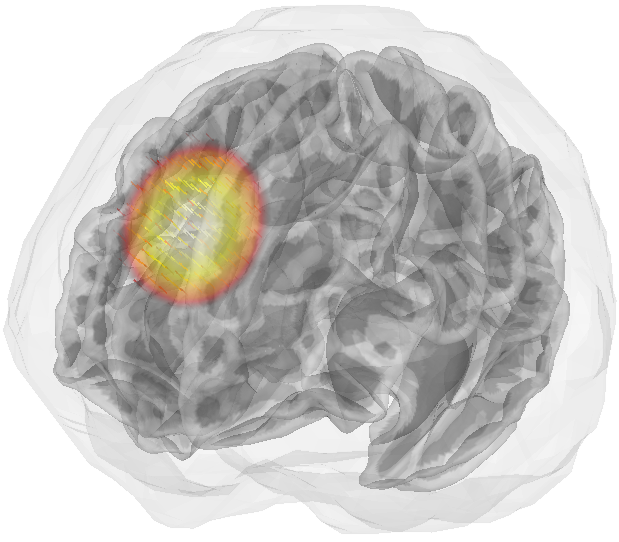}
    \end{subfigure}
    \hfill
    \begin{subfigure}[b]{0.21\textwidth}
        \centering
        \includegraphics[width=\textwidth]{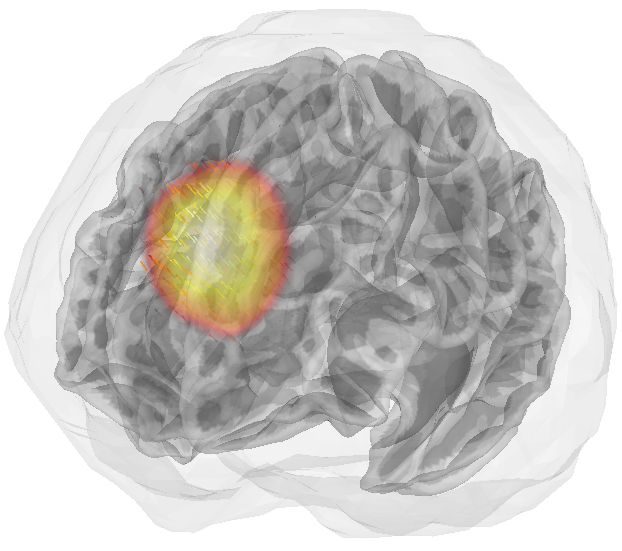}
    \end{subfigure}
    \hfill
    \begin{subfigure}[b]{0.21\textwidth}
        \centering
        \includegraphics[width=\textwidth]{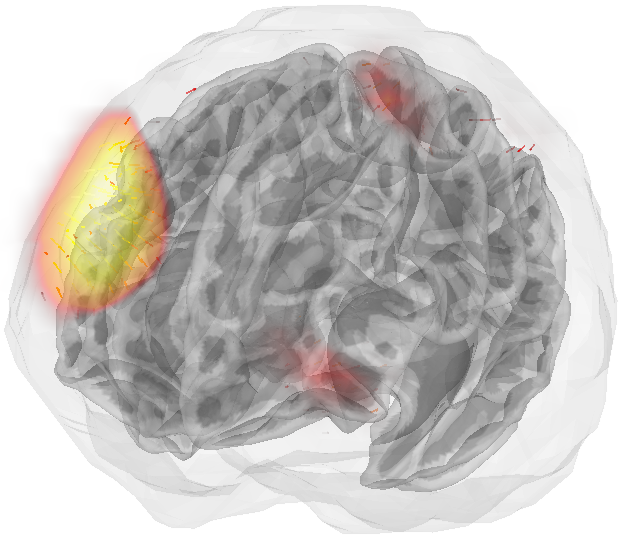}
    \end{subfigure}

    \begin{subfigure}[b]{0.21\textwidth}
        \centering
        \includegraphics[width=\textwidth]{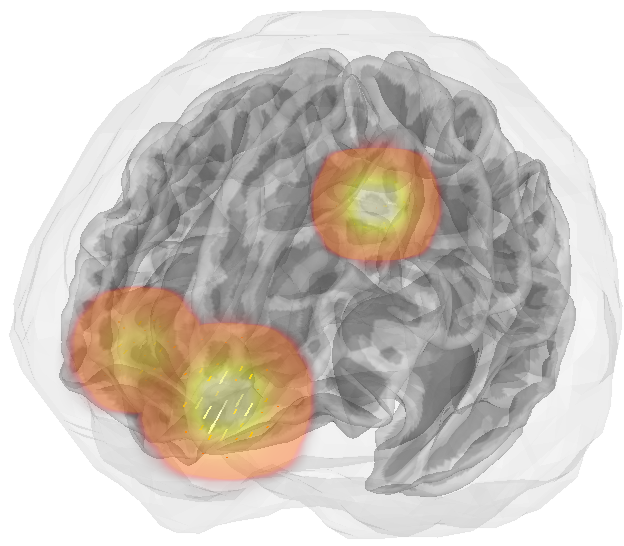}
        \caption{Ground Truth}
    \end{subfigure}
    \hfill
    \begin{subfigure}[b]{0.21\textwidth}
        \centering
        \includegraphics[width=\textwidth]{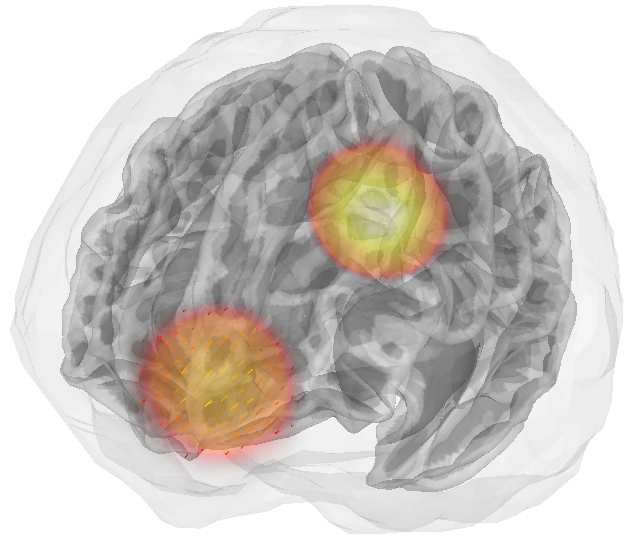}
        \caption{\mymethod}
    \end{subfigure}
    \hfill
    \begin{subfigure}[b]{0.21\textwidth}
        \centering
        \includegraphics[width=\textwidth]{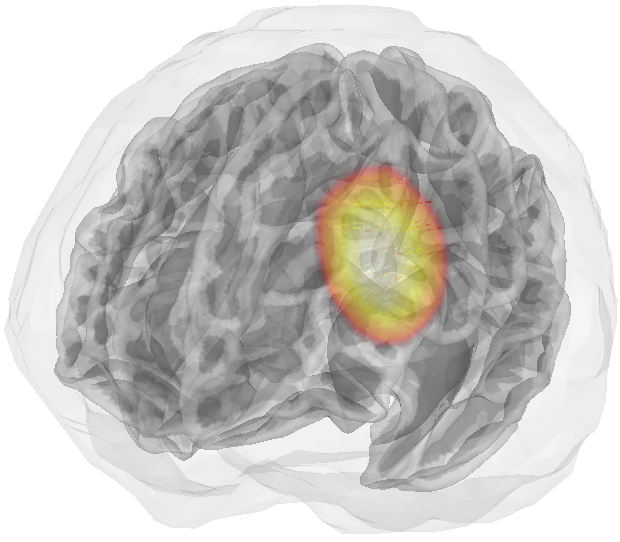}
        \caption{FCN}
    \end{subfigure}
    \hfill
    \begin{subfigure}[b]{0.21\textwidth}
        \centering
        \includegraphics[width=\textwidth]{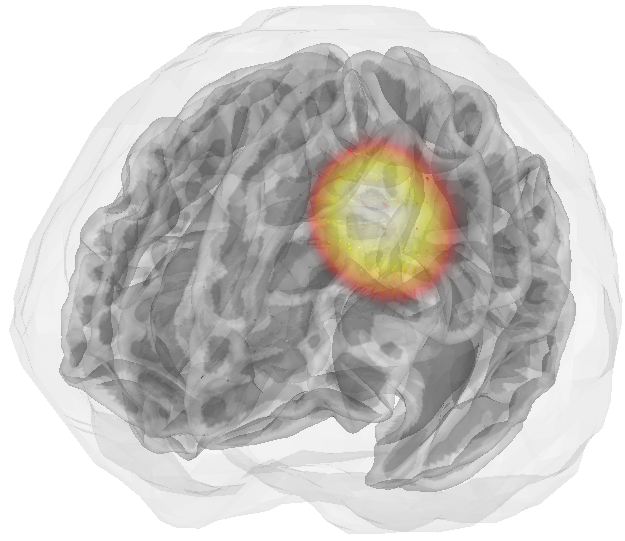}
        \caption{eLORETA}
    \end{subfigure}
    
    \caption{The ground truth activation and reconstruction from \mymethod~, the fully connected Network and eLORETA for \textbf{top} a single active source with SNR of $0$dB and \textbf{bottom} 3 active sources with SNR of $40$dB. \mymethod~incorporates the learned data prior and refines the shape of an active region to match the ground truth. When multiple sources are active, it does focus on the strongest sources, missing weak activations in other parts of the brain.}
    \label{fig:Visualsources}
\end{figure*}

\label{sec:metrics}
To evaluate the performance, we use the following three metrics, each investigating a different aspect of our prediction.
The Normalized Mean Squared Error compares the voxel wise prediction, considering both amplitude and orientation, while the Earth Mover Distance considers also spatial distances of the prediction. Lastly, as we consider free-orientated sources, we evaluate the angular accuracy of the predicted sources.
\subsubsection{Normalized Mean Squared Error}
Given the ground truth sources $\x \in \mathbb{R}^{3N}$ and predicted sources $\hat{\x} \in \mathbb{R}^{3N}$~and $\x_i \in  \mathbb{R}^3$ defined as the oriented activation of source $i$, the Normalized Mean Squared Error (NMSE) measures the voxel- and direction-wise deviation normalized by the total activation of the ground truth:
\begin{equation}
    \text{NMSE}(\x,\hat{\x}) = \frac{\sum_{i=0}^{N}  ||\x_i-\hat{\x}_i||_2^2}{\sum_{i=0}^{N}  ||\x_i||_2^2}
\end{equation}
By normalizing with the total activation per sample, we ensure that all samples contribute equally to the validation loss, regardless of their overall activation strength, while preserving relative source amplitudes within each sample.

\subsubsection{Normalized Earth Mover's Distance}
To incorporate the spatial arrangement of brain sources and penalize larger misplacement of activations for distributed dipoles, Haufe et al. \cite{haufe2008combining} proposed the use of the Earth Mover's Distance (EMD)~\cite{rubner2000earth}. It measures the minimal transport, based on the Euclidean distance in 3D space, needed to align the predicted and real source activations, thereby capturing the location error.
Since brain sources often consist of multiple, spatially extended activation areas, we compute EMD between the predicted $\hat{\x}$ and ground truth $\x$ source activations after normalizing them into probability distributions:
\begin{equation}
\x^A_i = \frac{||\x_i||_2}{\sum_j^N ||\x_j||_2}; \quad \hat{\x}^A_i = \frac{||\hat{\x}_i||_2}{\sum_j^N ||\hat{\x}_j||_2}
\end{equation}

The EMD is then computed as the solution to the following constrained optimization problem:
\begin{align}
\text{EMD}(\x,\hat{\x}) &= \min_{F} \sum_i^N \sum_j^N F_{ij}M_{ij} \\
\text{s.t. \qquad } F &\ge 0  ; \quad \sum_j^N F_{ij} = \x^A_i  ; \quad
\sum_i^N F_{ij} = \hat{\x}^A_j
\end{align}

where $M_{ij} = ||p_i-p_j||_2$ is the Euclidean distance between source locations $i$ and $j$ in 3D space. We solve this optimization problem using the publicly available POT package \cite{flamary2021pot}.
To unify different simulation settings, we propose to normalize the EMD by the EMD of the ground truth and an uniform distribution.
Specifically, we define the normalized EMD (NEMD) as follows:
\begin{equation}
\text{NEMD}(\x,\hat{\x}) = \frac{\text{EMD}(\x,\hat{\x}) }{\text{EMD}(\x,\textbf{u}) }
\end{equation}

where $\textbf{u} = \frac{1}{N}\mathbf{1}$ represents a uniform activation distributed equally across all potential source locations. This normalization step accounts for variations in source activation area due to multiple sources or large source extents, preventing automatic reductions in EMD caused by distributed activations rather than true mislocalization.

\subsubsection{Weighted Cosine Distance}
We consider free orientated sources and need to measure whether the models achieve good angular precision. Therefore, we compute the cosine distance at active source location between the ground truth and estimated activations.
\begin{equation}
    \text{WCos}(\x,\hat{\x}) = 1 - \sum_{i}
^{N} \x^A_i \frac{\x_i^\top \cdot \hat{\x}_i}{||\x_i||_2 ||\hat{\x_i}||_2}
\end{equation}
The relative amplitude $\x^A_i$ filters out non-active voxels and scales the importance proportional to the activation strength of the ground truth. 

\subsection{Baselines}
As a neural network competitor, we restrict our comparison to models that operate on a single time step, as our method does not leverage temporal context. Therefore, we consider the ConvDip model from \cite{hecker2021convdip} and a fully connected network (FCN) with 4 fully connected layers and ReLU activation, upscaling the design of previous work \cite{hecker2022long}. The first layer maps from the $61$ measurements to $C=1024$ hidden dimensions, followed by two hidden layers with the same dimensionality. The last layer maps from the $1024$ hidden dimensions to $4819 \times 3=14457$ output channel. Note that this simple architecture already has around $17$ million parameters, more than the $11$ million parameters of \mymethod. We investigate different architecture sizes in Section \ref{sec:computeefficiency}. 
As classical baselines, we use eLORETA \cite{pascual2007discrete} as a minimum norm solution. Additionally, we compare against Lasso \cite{tibshirani1996regression} with a group lasso penalty across three spatial dimensions, 
although we expect sparse methods to perform poorly with wide sources. Hyperparameters for these methods are optimized on the training data via a grid search.
We evaluate all approaches on the new head geometry, with a distinct number of possible source locations. Since ConvDip and FCN have a fixed output dimensionality, we morph their predictions to the new geometry using Symmetric Diffeomorphic Registration (SDR) \cite{avants2008symmetric}. In contrast, \mymethod, similar to the classical approaches eLORETA and Lasso, does not require SDR and instead adapts naturally by replacing the pseudo inverse. To ensure a realistic evaluation and avoid the inverse crime, this pseudo-inverse is computed using the conductivity values employed during training, rather than those used for data generation (see Sec.~\ref{sec:forwardmodeling}).

\section{Experiments}
\label{sec:experiments}
In this section, we evaluate \mymethod~on different synthetic settings to discover its advantages and limitations.
\subsection{Visual Results}

In Fig. \ref{fig:Visualsources}, we visually compare source predictions for eLORETA and \mymethod~for a single active source with 20mm extent reconstructed from single timestep EEG measurements for a single active source with $0$dB noise (top) and three active sources with $40$db noise (bottom). \mymethod~reconstruction demonstrates a clear refinement of the eLORETA solution, more accurately capturing the spatial extent and peak activation of the ground truth.  

\subsection{Effect of Signal to Noise Ratio}
\begin{figure*}[t]
    \centering
    \vspace{-0.2cm}
    \includegraphics[width=1.0\textwidth]{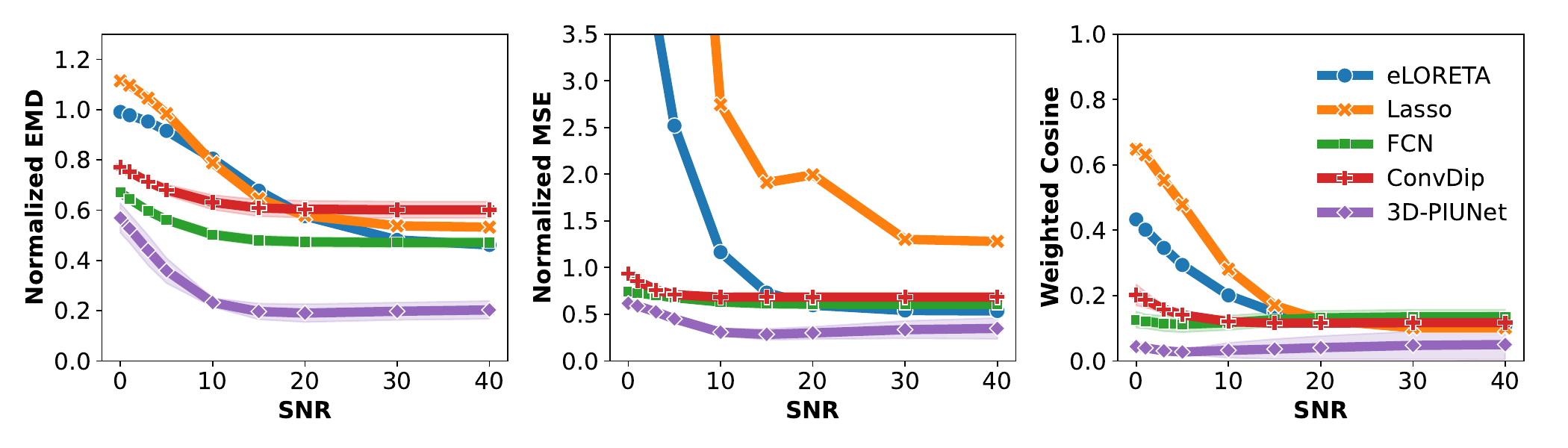}
    \caption{ Evaluation on the validation set across different Signal-to-Noise Ratios (SNR). The shaded area represents the standard deviations calculated from 5 trained models, emphasizing the consistency and reliability of the observed trends.
    \textbf{Left:} Normalized EMD decreases as SNR increases, with \mymethod~consistently showing the lowest values, indicating superior performance in minimizing distribution discrepancies.
    \textbf{Center:} For low SNR, the neural network approaches substantially outperform eLORETA. The discrepancy shrinks as SNR increases, with \mymethod~consistently achieving lower mean squared error. 
    \textbf{Right:} Weighted Cosine Distance remains low across all SNR levels for the neural networks, suggesting robust similarity in orientation, while eLORETA shows a significant improvement with increasing SNR.
    }
    \label{fig:SNRFigure}
\end{figure*}
In this experiment, we systematically vary the SNR from high noise setting with $0$dB to nearly noise-free observations of $40$dB, with the latter containing less noise than the maximum SNR of $30$dB seen during training. The results are depicted in Fig. \ref{fig:SNRFigure}.  We evaluated reconstruction performance using proposed metrics from section \ref{sec:metrics}: Earth Mover's Distance (EMD) to quantify differences in spatial distributions, Mean Squared Error (MSE) for overall amplitude accuracy, and weighted cosine similarity to assess orientation similarity. The neural network-based approaches (FCN, ConvDip and \mymethod) consistently outperform eLORETA and Lasso across all metrics and SNR levels with an exception of ConvDip for low noise scenarios. Since classical approaches reconstruct the measurements directly, they struggle with low SNR, while their performance converges toward that of neural networks as noise diminishes. Notably, \mymethod~exhibited superior performance compared to the FCN, particularly at SNR levels around 10 dB. This suggests that the U-Net architecture, with its ability to learn complex spatial patterns and non-linear relationships, is better equipped to leverage the additional information available in higher SNR data. The slight increase in variance observed at very high SNRs is likely due to the mismatch between the forward model used for data generation and the pseudo-inverse used for initialization. Nonetheless, the overall performance of \mymethod~remains consistently strong across all noise levels.

\subsection{Size of Sources}
\begin{figure*}[t]
    \centering
    \vspace{-0.2cm}
    \includegraphics[width=1.0\textwidth]{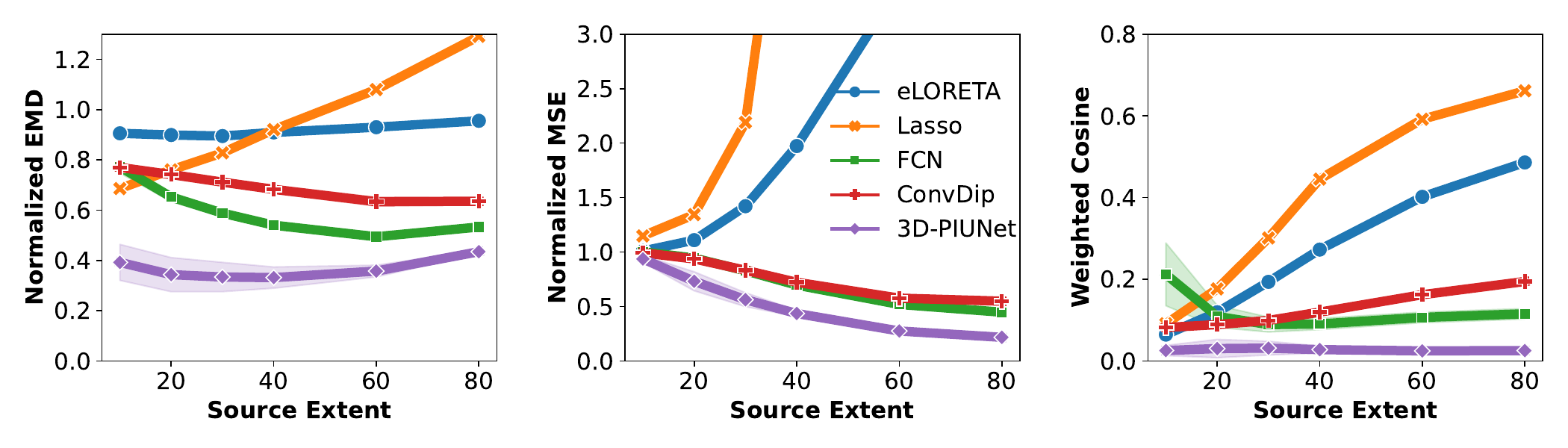}
    \caption{\mymethod~outperforms all baselines for different spatial extent of a single source with $5$dB noise. }
    \label{fig:SizeSources}
\end{figure*}
In Fig. \ref{fig:SizeSources}, we show model performance on different-sized sources for a single active source with an SNR of $5$dB. \mymethod~improves on the solution of eLORETA and outperforms the FCN with respect to all metrics. For small active sources with an extent of 10mm, we observe a low EMD, indicating that \mymethod~predicts activation close to the correct source, even outperforming the sparse method Lasso, which suffers from the small SNR as shown in the previous section. However, the high Normalized MSE suggests that none of the approaches perfectly matches the source on a voxel level. This trade-off may arise from the model's focus on accurately localizing sources, even when their exact amplitude is harder to capture at this resolution. Notably, \mymethod~demonstrates robustness to variations in source extent, effectively handling both sparse and spread sources despite being trained on a general dataset. This suggests the model has learned a versatile representation of source activity patterns.

\subsection{Number of Sources}
\begin{figure*}[t]
    \centering
    \vspace{-0.2cm}
    \includegraphics[width=1.0\textwidth]{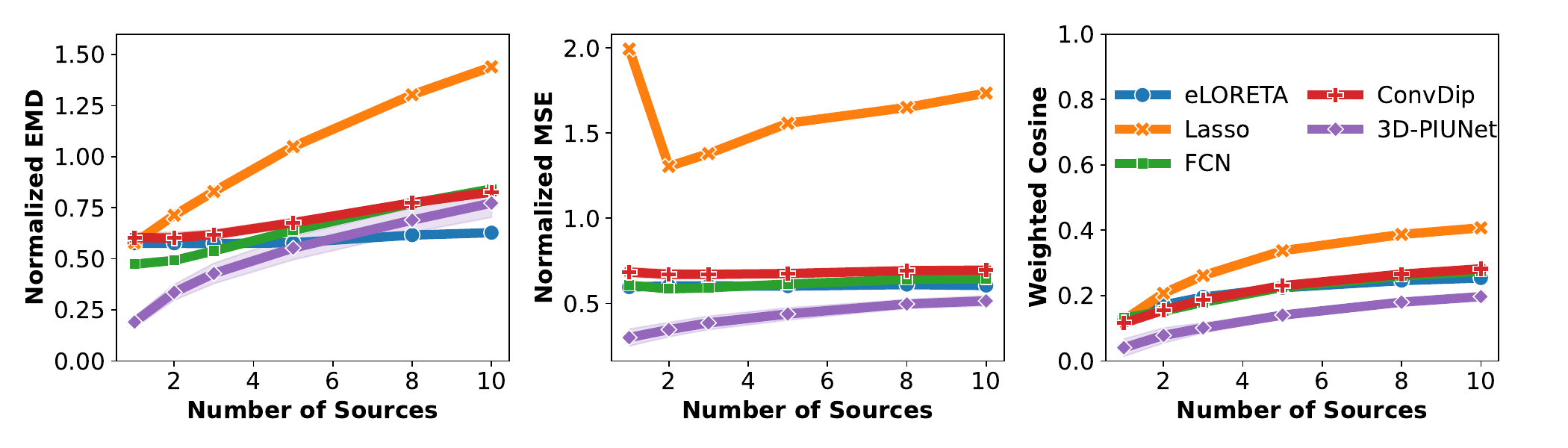}
    \caption{We show the performance of increasing number of sources for $20$dB Noise with varying extension. Even though the performance drops with more than 3 active regions, \mymethod~remains superior to the other methods for up to 5 active sources. }
    \label{fig:NumberSources}
\end{figure*}
In Fig. \ref{fig:NumberSources}, we analyze model performance for different numbers of active sources for the medium noise setting of 20dB and sources with random extent between $10$ and $80$mm. \mymethod~consistently outperforms the FCN across all metrics, but shows a decrease in performance as the number of sources increases, particularly in terms of the normalized EMD. 
Interestingly, the performance of eLORETA remains relatively constant regardless of the number of active sources. This is likely because eLORETA is a linear method that distributes the estimated source activity smoothly across the entire source space. As the number of sources increases, eLORETA continues to spread the activity, leading to a consistent but less accurate reconstruction compared to the neural network models, which can learn more complex, non-linear relationships between sources and sensor data. 
As we consider only a single time step the signal of the sources overlap and the demixing becomes a very hard task, even when a data prior is learned. In future work, we aim to add the time dimension with varying source amplitudes, providing additional information necessary for separating the different regions. 

\subsection{Transferability to Real Data}

\begin{figure*}[h!]
    \centering
    \begin{subfigure}[b]{0.49\textwidth}
        \centering
        \includegraphics[width=\textwidth]{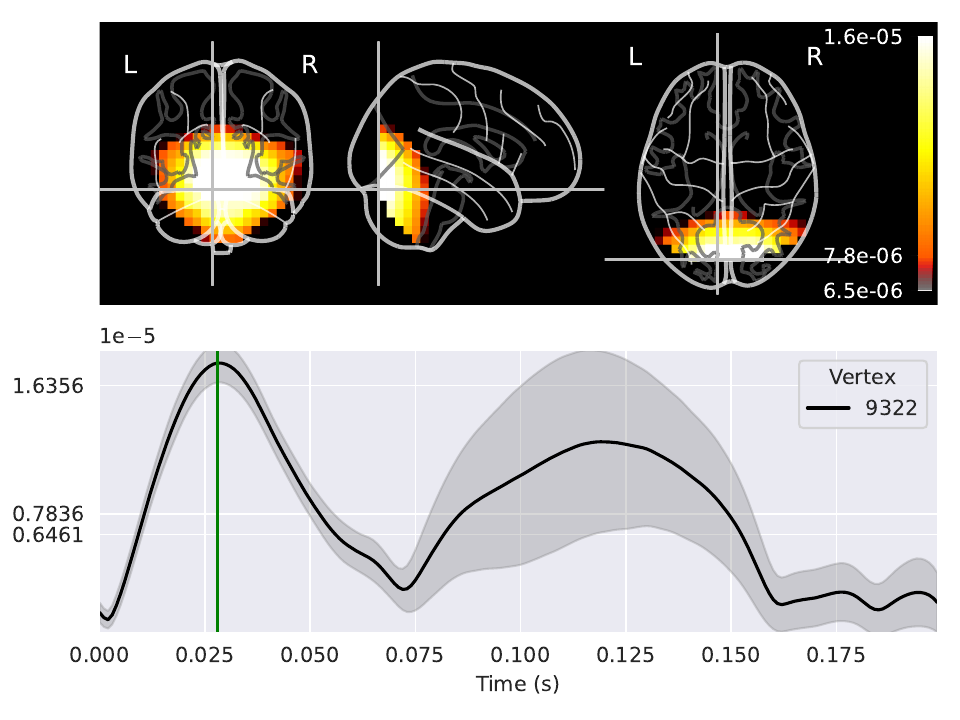}
        \caption{eLORETA}
        \label{fig:figure1}
    \end{subfigure}
    \hfill
    \begin{subfigure}[b]{0.49\textwidth}
        \centering
        \includegraphics[width=\textwidth]{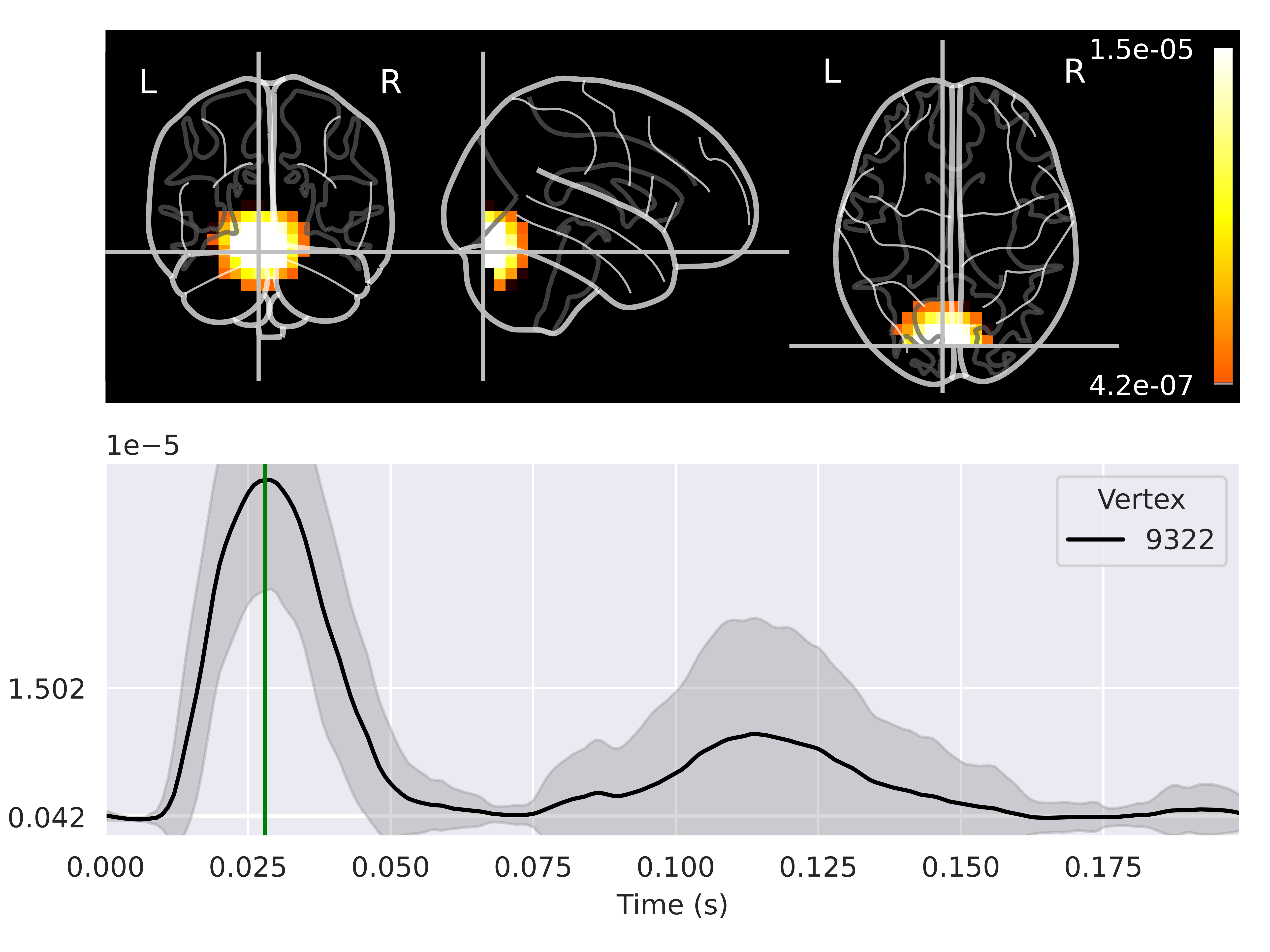}
        \caption{\mymethod}
        \label{fig:figure2}
    \end{subfigure}
    \caption{This figure illustrates the average predicted source activation for 20 ensembled stimuli from the THINGS-EEG dataset. Both panels display activation peaks of approximately 30 ms and around 110 ms post-stimulus exposure. In panel (a), the eLORETA solution shows a diffuse activation pattern, indicating a broader but less specific response within the visual cortex. In contrast, panel (b) demonstrates that \mymethod~predicts localized and concentrated activation, specifically highlighting the visual cortex region. 
    }
    \label{fig:thingseegpred}
\end{figure*}
Using the pretrained model from the last sections, we validate \mymethod~on real data, utilizing the THINGS-EEG2 dataset \cite{gifford2022large}. This large-scale visual dataset followed the Rapid Serial Visual Presentation (RSVP) paradigm \cite{intraub1981rapid,grootswagers2019representational} where images are displayed for 100 ms each, followed by a 100 ms blank screen in a series of 20 images. As the full dataset is large, we focus on the test set of subject 10, containing 20 trials of EEG recording for each image stimulus. We applied multivariate noise normalization \cite{guggenmos2018multivariate}, following the original work. 

The THINGS-EEG2 dataset does not include MRI data for the subjects. Consequently, we use the same head geometry from section \ref{sec:forwardmodeling} and recomputing the forward and pseudo-inverse for the new sensor layout.~
To enhance the signal-to-noise ratio, we apply ensemble averaging \cite{SORNMO2005181}, combining EEG signals from $20$ instances where the same image was presented to the subject. We apply eLORETA and \mymethod, to the preprocessed EEG data to estimate the source activations. As both are only spatial models, we predict each time step independently. 
Note that \mymethod~is trained only on the artificial data and applied to the different sensor layout without any fine-tuning by replacing just the pseudo inverse. Classical end-to-end approaches would require at least retraining of the input layer to account for the changed input dimensionality. 

The predicted source activation averaged over the 200 test image conditions is shown in Fig. \ref{fig:thingseegpred}. We observe two main activations in the visual cortex, the first around 30~ms and the second 110~ms after stimulus exposure. The first sharper peak after 25ms appears unusually early. As the experiments follow the RSVP paradigm, we assume that this first peak corresponds to the blank screen, which triggers a more uniform brain response.
The second peak aligns well with canonical visual evoked responses, typically occurring between 75-140ms \cite{creel2019visually}, and is consistent with the image-related cortical activity reported by Gifford et al. \cite{gifford2022large}.

The spatial characteristics of the activations differ significantly between the two methods.
For eLORETA, the spatial activation pattern appears diffuse, while \mymethod~is more localized, precisely highlighting the visual cortex. This concentrated activation suggests that \mymethod~is better at pinpointing specific regions of neural activity. The temporal activation is less pronounced but displays a similar pattern to eLORETA. This verifies the effectiveness of the proposed method also for real-world data. 

\subsection{Computational Efficiency}
\label{sec:computeefficiency}

\begin{table}[t]
    \centering
    \caption{Comparison of different model configurations with respect to runtime and EMD for 2000 single source activations with $20$ dB SNR and varying source extent. While \mymethod~is slower than the Fully Connected Network, its runtime is still sufficiently fast for real time analysis. The number of parameters is given in million.
    }
    \begin{tabular}{l|cccccc}
        \toprule
        \textbf{Model} & \textbf{Depth} & \textbf{Channels} $C$ & \textbf{Parameter} & \textbf{Runtime} & \textbf{EMD} \\
        \midrule
        eLORETA & - & - & 0.0M & 0.003s & 0.58 \\
        Lasso & - & - & 0.0M & 8.178s & 0.72 \\
        \midrule
        FCN & 3 & 512  & 8.0M & 0.006s & 0.52 \\
        FCN & 3 & 1024 & 17.0M & 0.006s & 0.47 \\
        FCN & 4 & 1024 & 18.0M & 0.007s & 0.49 \\
        FCN & 3 & 4096 & 93.0M & 0.006s & 0.40 \\
        FCN & 3 & 16384 & 774.0M & 0.007s & 0.38 \\
        ConvDip & 3 & 512 & 8.0M & 0.633s & 0.60 \\ 
        \midrule
        3D-PIUNet & 3 & 8 & 0.7M & 0.967s & 0.21 \\
        \textbf{3D-PIUNet} & 3 & 32 & 11.1M & 3.234s & 0.20 \\
        3D-PIUNet & 4 & 32 & 19.1M & 3.230s & 0.19 \\
        3D-PIUNet & 3 & 64 & 44.2M & 6.721s & 0.20 \\
        \bottomrule
    \end{tabular}
    
    \label{tab:model_comparison}
\end{table}

While \mymethod~offers superior reconstruction accuracy, it is important to consider its computational demands. Training \mymethod~requires approximately 5 hours on a single A100 GPU, compared to 2 hours for the FCN. However, the primary concern for practical applications is inference time. In our evaluation on a validation set of 2000 samples (Table \ref{tab:model_comparison}), both eLORETA and the Fully Connected Network perform inference for 2000 samples in a fraction of seconds. \mymethod, while slower, still completed inference a few seconds, notably faster than Lasso. 
To further examine model efficiency, we compared different hyperparameter settings. While scaling up the FCN improves its performance, there remains a big gap even to the smallest version of \mymethod, demonstrating the parameter efficiency of the convolutional network. For \mymethod, we varied not only the number of channels $C$, but also the depth, defined as the number of encoder-decoder resolution levels (i.e., downscaling and upscaling stages). 
Overall, \mymethod~shows stable performance across different hyperparameter settings.

\subsection{Loss function Overview}
\label{sec:lossFunctions}
\begin{figure}
    \centering
    \vspace{-0.1cm}
    \includegraphics[width=\linewidth]{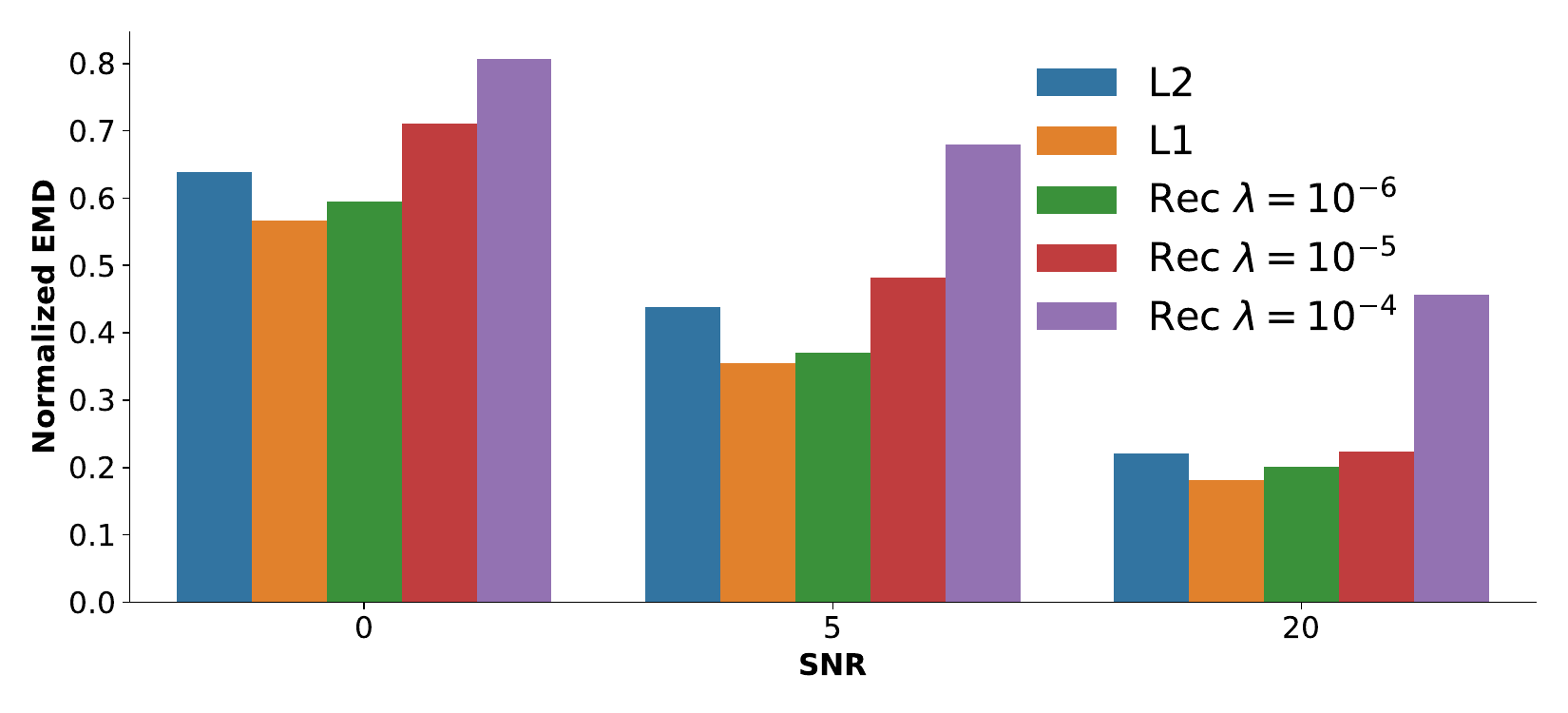}
    \vspace{-0.1cm}
    \caption{ Comparison of training \mymethod~with different loss functions. The proposed $\ell_1$ loss outperforms the $\ell_2$ loss for a single active source. Adding a data reconstruction loss does harm the performance.}
    \label{fig:LossOverviewMain}
\end{figure}
In this section, we evaluate the proposed \mymethod~trained with both $\ell_2$ and a data reconstruction loss on single-source prediction across three different noise levels, as shown in Fig. \ref{fig:LossOverviewMain}. The reconstruction loss is based on the forward model and transfers the prediction back to  measurement space: 
\begin{equation}
    \mathcal{L}_{rec}(\x, \hat{\x},\y) = ||\hat{\x} - \x||_1 + \lambda||\A \hat{\x} - \y||_2^2
\end{equation}
However, due to the noisy measurements, this reconstruction loss drives the network's predictions away from the ground truth, degrading performance. Overall, our results indicate that the $\ell_1$ loss in source space is the most effective for training the neural network.

\section{Discussion \& Conclusion}
Inverse problems such as source localization are notoriously hard to solve due to their ill-posed nature. While end-to-end learning models such as neural networks have contributed to the recent progress in the field, we propose the hybrid approach, \mymethod, which combines the strengths of physics-based modeling and deep learning. 
\mymethod~combines a not-learned physics-based pseudo-inverse with a learned 3D convolutional network operating directly in source space. This synergy achieves significant performance gains, even when using a standard head model. Our results demonstrate enhanced reconstruction accuracy and noise robustness compared to current state-of-the-art methods, effectively predicting both sparse and extended sources without requiring specific training for different types. This advancement could provide deeper insights into brain function and pave the way for improvements in clinical applications, such as diagnosing neurological disorders. 

\subsection{Limitations \& Future Work}

The effectiveness of our method depends on the quality of the data prior, which is shaped by the simulations used during training. While we employed a diverse set of simulations, varying the size and number of sources, our approach assumes a Gaussian prior on the source shape. This assumption, while effective for controlled evaluations, may not fully capture the diversity and complexity of real neural activations, potentially limiting performance in highly heterogeneous clinical datasets. Future work could explore more biologically informed priors, such as those based on functional parcellations \cite{eickhoff2018imaging}, to enhance realism. Additionally, incorporating known correlation structures between sources, as common in neural mass models \cite{wilson1972excitatory,cakan2021}, could further refine the learned prior and improve source reconstruction in real-world settings.

Furthermore, our focus on the spatial aspects of source localization neglects temporal dynamics. Given the inherently ill-posed nature of EEG/MEG source localization, using temporal information could enhance accuracy, particularly in resolving multiple simultaneously active sources. However, the proposed hybrid approach is general and the core idea of combining pseudo-inverse with deep learning can be adapted to also include the time domain,  which is a direction we plan to pursue in future work. This extension is particularly relevant for clinical applications that focus on the temporal dynamics of neural events, such as brain-computer interfaces.

Like many advanced learning approaches, the proposed \mymethod~presents challenges in interpretability. Neural networks, while highly effective, can sometimes be seen as ``black boxes''. This can pose challenges in clinical adoption, where transparency and trust in model decisions are essential. In future research, we aim to integrate the recent advances on neural network explainability \cite{samek2021explaining} to improve the transparency and reliability of neural network-based source localization methods.

\bibliographystyle{IEEEtran}
\bibliography{main}


\newpage
\onecolumn

\appendix
\subsection{Different Pseudo Initialization}
\begin{figure*}[h]
    \centering
    \vspace{-0.1cm}
    \includegraphics[width=\textwidth]{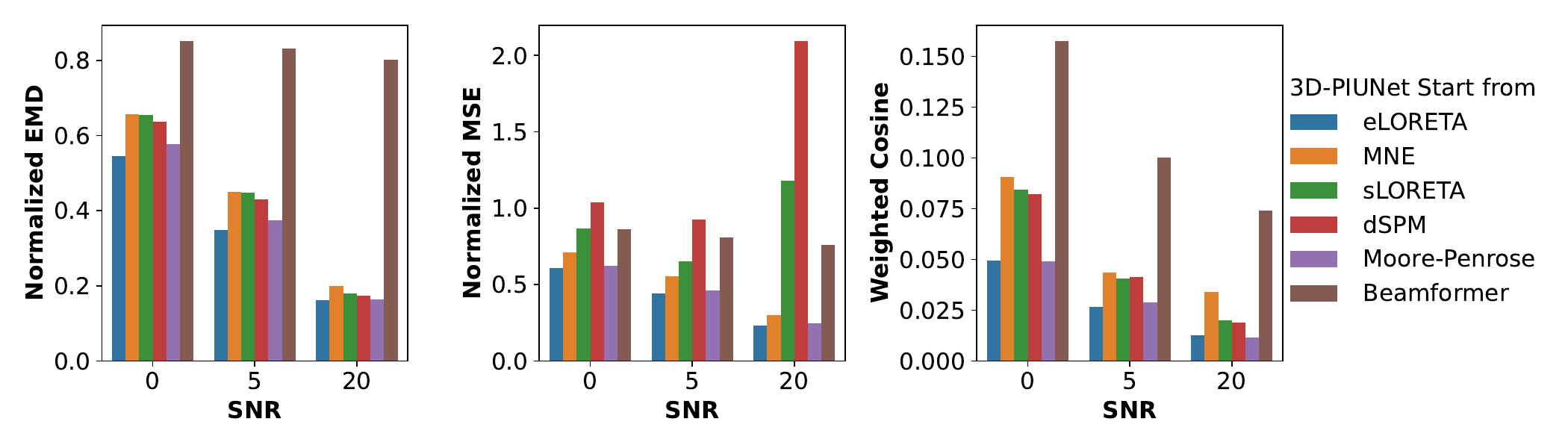}
    \vspace{-0.1cm}
    \caption{Model performance when training the 3D Convolutional Network starting from different Pseudo Inverse solutions.}
    \label{fig:PseudoInitialization}
\end{figure*}
\label{sec:pseudoInitVariation}
In this section, we investigate the relevance of the pseudo-inverse solution for model performance by training a 3D convolutional network with different pseudo-inverse solutions as starting points. We compare the proposed eLORETA pseudo-inverse with minimum norm estimates (MNE), sLORETA, dSPM, and beamformer solutions implemented in the MNE package \cite{GramfortEtAl2014}, as well as the Moore-Penrose inverse. Fig. \ref{fig:PseudoInitialization} illustrates performance on single-source prediction across three different noise levels. While the performance of the regularized minimum norm solutions is similar, the eLORETA solution achieves the best results. The beamformer, however, shows reduced performance, likely due to its sensitivity to noise covariance estimation and source correlation, which can lead to suboptimal localization and model performance under varying SNR conditions. Overall, \mymethod~demonstrates robust performance across different pseudo-inverse initializations, highlighting its adaptability and effectiveness regardless of the chosen inverse method.

\subsection{Robustness to Missing Sensors}

\begin{figure*}[h]
    \includegraphics[width=\textwidth]{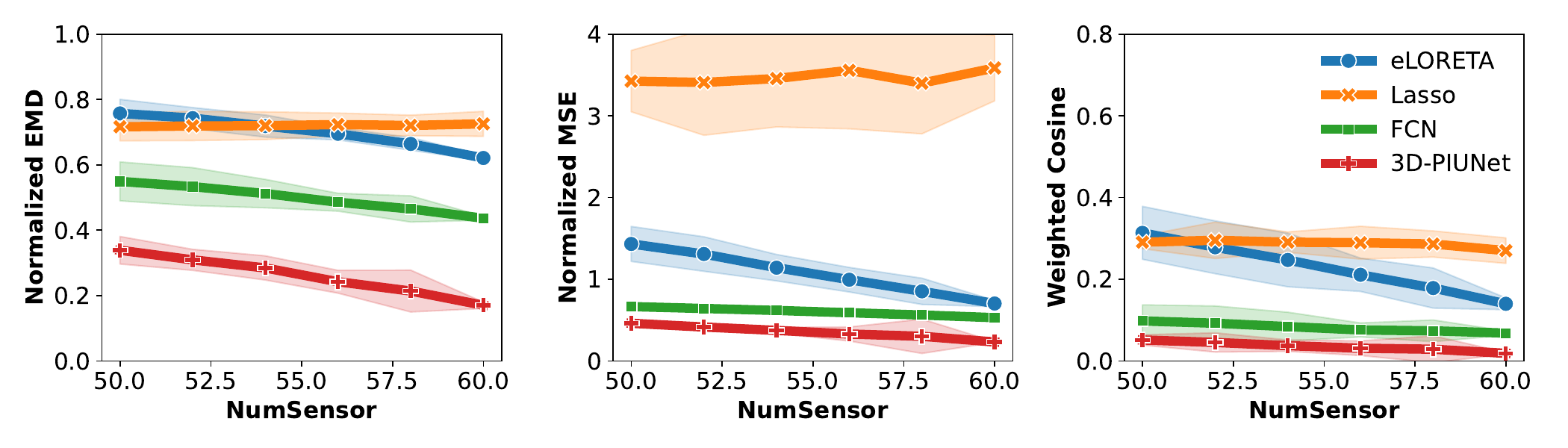}
    \caption{We evaluate the model performance for a reduced set of sensors, by using only a subset of the 61 available sensors. While there is a small decrease in performance, \mymethod still performs best, demonstrating robustness to missing values.}
    \label{fig:NSensors}
\end{figure*}

\label{sec:missingSensors}
In this section, we evaluate the model's robustness to missing sensor data. Starting with the original 61 sensors, we randomly set up to 11 sensor measurements and their corresponding entries in the lead field matrix to zero, simulating missing data. Given that the impact of missing sensors can vary depending on their locations, we plot the standard deviation of performance across five different random subsets of missing sensors. The results for a single active source with an SNR of $20$dB are shown in Fig. \ref{fig:NSensors}. Both neural network approaches demonstrate greater robustness to missing sensor values compared to eLORETA, suggesting that the learned data priors effectively compensate for measurement errors and improve resilience to data loss.

\end{document}